\documentstyle[editedvolume,epsfig,numreferences]{crckapb} 

\newcommand{\sam}{{\sc sam}}
\newcommand{\w}{\cal W}
\newcommand{\la}{\left <}
\newcommand{\ra}{\right >}

\begin{opening}
\title{A SCALAR ARCHING MODEL}
\author{P. CLAUDIN}
\author{J.-P. BOUCHAUD}
\institute{Service de Physique de l'Etat Condens\'e,\\
CEA, Orme des Merisiers,\\
91191 Gif-sur-Yvette, Cedex France.}
\end{opening}
\runningtitle{A SCALAR ARCHING MODEL}

\begin{document}

\begin{abstract}
In this paper we propose an extension of the model of
Liu, Coppersmith et al. \cite{Liu,Copper}, in order to allow for arch formation.
This extended model qualitatively captures interesting properties of granular materials
due to fluctuations of stress paths, such as stress fluctuations and stick-slip
motion.
\end{abstract}

\section{Introduction}

Granular media are materials where fluctuations are very large.
When filling a silo with grains, a part of the weight of the
grains is supported by the walls of the silo, meaning that the bottom
plate of the silo only carries an {\it apparent weight} $\w$.
This effect is well known and was studied by Janssen at the end of
the last century.
When repeating this procedure with the very same amount
of grain, one observe large fluctuations of $\w$, typically
of order $10 - 25 \%$! On a given silo, $\w$ can also vary drastically
because of very small perturbations, such as variations
of temperature or density \cite{Vanel,Loggia}.

These effects can be understood in terms of {\it arching}. As a matter
of fact, stress propagation in granular media is strongly inhomogeneous:
clear stress paths are present and carry an important part of the total
weight of the grains. Those paths, or arches, are completely different
from a realisation to another, leading (potentially) to very different
values of $\w$. Furthermore, the geometry of these paths can easily rearrange
under small perturbations, inducing strong fluctuations in  $\w$.

Another phenomenon closely connected to the presence of these arches
is that of the stick-slip motion. Imagine  trying to push some granular
material through a tube with a piston. Moments where the system jams (stick)
 and moments where the system slides
(slip) will irregulary alternate. This is explained by the fact that the stress paths
network is sometimes able to resist to the applied force, and sometimes
is not, depending on its structure (which fluctuates).

In this paper, we propose a very simple numerical model which
qualitatively captures some features of the above effects.
This model will be called the Scalar Arching Model, or \sam\
in the following.

\section{The \sam}

The model we are going to present is an extension of that proposed
by Liu, Coppersmith et al. \cite{Liu,Copper}, which
allows for arch formation. This model only deals with the vertical
normal component of the stress tensor, $\sigma_{zz} = w$, the `weight'.
In that sense, the model is {\it scalar}.
This is obviously a simplifying choice which may not be justified.
Work is in progress for extending this model to a fully tensorial description
(see e.g. \cite{us}). Again for simplicity, we will keep to a two dimensional situation. Note that arching effects are expected to be
more pronounced than in three dimensions.
\begin{figure}[htb]
\begin{center}
\epsfysize=5.5cm
\epsfbox{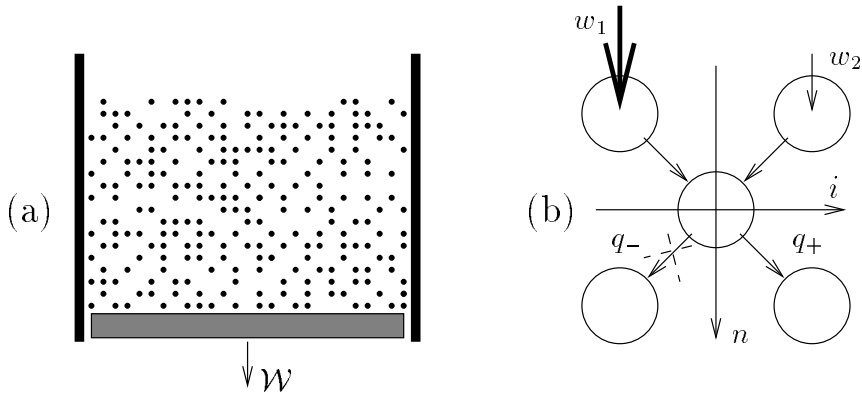}
\caption{Figure (a) represents the system we study. An apparent weight
$\w$ is measured at the bottom plate of a silo. Figure (b) shows the
underlying lattice with which we modelize the granular medium. Indexes
$i$ and $n$ are such that $-L \le i \le L$ and $0 \le n \le H$, where
$2L$ and $H$ are respectively the width and the height of the silo.
On the particular configuration of figure (b), the force $w_1$ is much
larger than $w_2$, meaning that the shear force acting on the grain (i,n)
is strong enough to remove the contacts between the grains $(i,n)$ and
$(i-1,n+1)$.
\label{shema}}
\end{center}
\end{figure}
The system we would like to describe is shown in figure \ref{shema}-(a).
Following Liu et al.'s approach, the granular packing is represented by
a regular square lattice. Each site is a `grain' labelled by two
integers $(i,n)$ giving its horizontal and vertical coordinate.
All the randomness of the local packing, the friction, the size and
shape of the grains, is assumed to be encoded into random transmission coefficients $q_\pm$. The rule for the propagation of the weight is the
following:
\begin{equation}
w(i,n)=w_g + q_+(i-1,n-1)w(i-1,n-1) + q_-(i+1,n-1)w(i+1,n-1)
\label{propa}
\end{equation}
where $w_g$ is the weight of a grain. This rule simply means that each
grain supports the force of its two upwards neighbours and share its
own load randomly between its two downwards neighbours. (The r\^ole of correlations has been recently discussed in \cite{Mario}). The mass
conservation constraint imposes $q_+(i,n) + q_-(i,n) = 1$. At this stage,
the model is the one considered in \cite{Liu,Copper}.
We now include a {\it local slip condition}: when the shear on a given
grain is too strong, the grain can slip and lose its contact with its
neighbours opposite to the direction of the shear. More precisely,
one defines a threshold $R_c$ such that
\begin{equation}
q_\pm(i,n) = 1-q_\mp(i,n)=0 \qquad  \mbox{if} \qquad
{w_\mp-w_\pm} \ge R_c {w(i,n)}\label{lsc1}
\end{equation}
where $w_\pm = q_\pm(i\mp 1,n-1) w(i\mp 1,n-1)$. These rules lead to
an avalanche-like process: suppose, as shown on figure \ref{shema}-(b),
that the link between the grains $(i,n)$ and $(i-1,n+1)$ is removed
because the force $w_1$ is much larger than $w_2$. The force from
$(i,n)$ to $(i+1,n+1)$ is then very likely to be much larger than the
force from $(i+2,n)$ to $(i+1,n+1)$, and the link between the grains
$(i+1,n+1)$ and $(i,n+2)$ is very likely to be removed as well. This
process (that was called {\it static avalanche} in \cite{SAM}) can be
interpreted as an {\it arch formation}.

A particular value of $\w$ is associated to a particular configuration
of the stress paths. At the walls, forces are absorbed with the following
rule: if $i=\pm L$, the fraction $q_\pm (\pm L,n)$ of the load $w(\pm L,n)$
is absorbed. Depending on the configuration of the stress paths, a larger
or smaller part of the weight of the grains is then supported by the walls,
leading to differents values of $\w$. We studied the probability
distribution function (p.d.f.) of the values of $\w$ from many different
configurations of the stress paths (i.e. from many different silos). The relative standard deviation from the mean value is as large as $20 \%$, as found experimentally. We also looked
at the p.d.f. of the local weight $w(i,H)$
and found it to be a power-law, meaning that extremely large
values are expected.

\begin{figure}[t]
\begin{center}
\epsfysize=6.5cm
\epsfbox{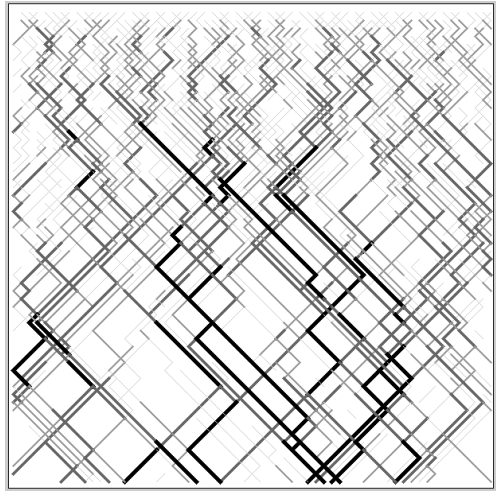}
\caption{This figure represents a particular configuration of the
stress paths obtained in a silo with the \sam. Lines are all the bolder
as the stress is larger.
\label{reseau}}
\end{center}
\end{figure}
The most striking feature of this model is that the stress paths network
generated by the \sam\ (as shown in figure \ref{reseau}) is very likely to
rearrange under small perturbations. Our control parameter for
generating perturbations is the threshold $R_c$. This coefficient does
not actually represent a specific physical quantity but can be seen
as representing the temperature and/or the compacity. As a matter of fact,
contacts between grains ar very sensible to small changes of those two
quantities. What we observed in the \sam\ is that when $R_c$ is changed
by a very small amount, the stress paths network
sometimes rearranges and sometimes does not. When it does, $\w$ changes
by a relative amount $\Delta$ which can be either small or large, meaning
that the actual value of $\Delta$ is not correlated to the amount by
which $R_c$ has been changed. Furthermore, $\Delta$ is found to be
power-law distributed: $P(\Delta) \sim 1/\Delta$, which means that large rearrangements of the stress paths network are as probable as small ones.
In that sense, the \sam\ captures the feature that granular matter is
extremely `fragile', i.e., sensitive to small perturbations.
Actually, our model is close in spirit to the large class of `SOC' (Self-Organized Critical)
models \cite{BTW}, initially proposed to describe true `dynamical'
avalanches in granular media.

\section{Stick-slip motion}

\begin{figure}[htb]
\begin{center}
\epsfysize=6.5cm
\epsfbox{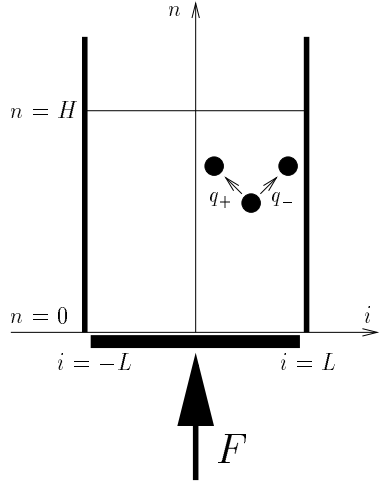}
\caption{The \sam\ can be slightly modified to describe the situation
where an upwards force $F$ is applied on the bottom plate of the silo.
Four parameters control the system: the aspect ratio $b$, $R_c$ the threshold of
the \sam , the `jamming ability' of the walls $\alpha$ and $p$ the rearrangement
probability (see below in the text).
\label{shema2}}
\end{center}
\end{figure}

When the bottom plate of the silo is used as an upwards pushing piston
(see figure \ref{shema2}), an irregular stick-slip motion of the system
is observed \cite{Kolb}.
This behaviour can be understood by the fact that the stress paths network
rearranges, generating configurations where it can resist to $F$
(sticking or jamming situations) and configurations
where it cannot (slipping or sliding situations). In this section, we
will explain how one can slightly modify the original version of the \sam\ 
to describe this situation.

The idea is to look at the \sam\ picture `upside down'. We neglect the weight of the
grains and focus on the transmission of $F$ through the grains,
from the bottom of the cell to the walls and the free surface.
All propagation rules are kept the same.
The only noticeable change from last section is what
happens at the walls. Because strong arching is expected at the walls when
$F$ is applied, we introduce a new parameter $\alpha$, the `jamming ability' of
the walls. With probability $1-\alpha$ absorbing rules of the last section
apply. With probability $\alpha$ however, the load $w(\pm ,L)$ is
completely absorbed by the wall, meaning that a local arch is strongly
`anchored' on the wall. Such situations are essential to get the system
jammed. In addition to the $q_\pm (i,n)$, we now then have new random
numbers $\alpha_\pm (n)$ which, compared to $\alpha$, determines which absorbing
rule applied at site $(\pm L,n)$.

We then propose the following algorithm. For a given force $F$, and a
given set of random numbers $q_\pm (i,n)$ and $\alpha_\pm (n)$,\\
$\bullet$ we calculate $F_w$ and $F_s$, respectively the total forces on the
walls and on the top surface of the silo. Obviously, $F=F_w+F_s$.\\
$\bullet$ if $F_s = 0$, the grains do not move. It is a {\it stick} situation.
We then increase the applied force $F$ of $\Delta F$ and the time $t$
of $\Delta t$. In order to mimick the mechanical noise, we also change all random numbers with probability $p$, and go back to the first point.\\
$\bullet$ if $F_s > 0$, the static equilibrium conditions for the system are not
satisfied, which means that grains are moving. It is a {\it slip} situation.
We then decrease the applied force $F$ by $\Delta F$, change all random
numbers (because the flow motion completely rearranges the packing), and go back to the first point.\\
The simulation starts at $t=0$ with $F=0$ and let $F(t)$ evolve.
It is important to note that our model is a pure static model: no dynamics
is included. Therefore, the motion of the grains during the slipping
moment is assumed to be infinitely quick.
We then actually describe only sticking situations, separated by slipping
moments which have two effects: untighten the spring governing the external
force $F$, and reinitialize the structure of the packing (i.e. the random
numbers). Such an algorithm leads indeed to an irregular stick-slip motion.

\begin{figure}[htb]
\begin{center}
\epsfysize=5cm
\epsfbox{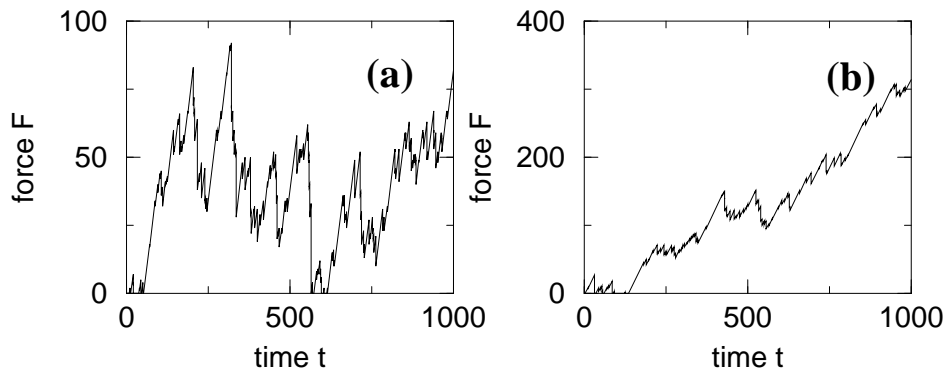}
\caption{These figures show typical curves of $F(t)$ in the
sliping stick-slip phase (a) and the sticking stick-slip phase (b).
\label{phases}}
\end{center}
\end{figure}
Depending on the values of the parameters controlling the system,
two different behaviours are observed.
For small values of the aspect ratio $b$ or the `jamming ability' $\alpha$,
situations where the system never jams for ever are typically observed,
see figure \ref{phases}-(a). We called these situations {\it slipping}
stick-slip phases. On the contrary, {\it sticking} stick-slip phases
are seen for large values of $b$ and $\alpha$, see figure
\ref{phases}-(b). The critical curves $\alpha_c (p)$ or $\alpha_c (R_c)$ can be
plotted for a fixed $b$, which separate the two regimes.
We caracterized the first phase by the first return time $\tau$, i.e. the
interval of time between two consecutive times where $F$ vanishes,
and the second one by the slope $s=F(t)/t$. Just below the critical point,
$\tau$ is be power-law distributed with an exponent $3/2$, meaning that
$F$ simply behaves like a random walk. When $\alpha \to \alpha_c$ we 
find $\la \tau \ra \sim 1/(\alpha_c-\alpha)$ and
$\la s \ra \sim \alpha-\alpha_c$ for $\alpha > \alpha_c$. More details can be found in \cite{stickslip}.

\section{Acknowledgements} We are grateful to E. Kolb, T. Mazozi, J. Duran,
E. Cl\'ement, J. Rajchenbach, D. Loggia and P. Mills for many fruitful
discussions on this model.

\end{document}